# $KTaO_3$-based editable superconducting diode

Yishuai Wang[1#], Wenze Pan[1#], Meng Zhang[1], and Yanwu Xie[1,2,3,4*]

[1]School of Physics, and State Key Laboratory for Extreme Photonics and Instrumentation, Zhejiang University, Hangzhou 310027, China

[2]College of Optical Science and Engineering, Zhejiang University, Hangzhou 310027, China

[3]Hefei National Laboratory; Hefei 230088, China.

[4]Collaborative Innovation Center of Advanced Microstructures, Nanjing University; Nanjing 210093, China

[#]These authors contributed equally to this work.

[*]To whom correspondence should be addressed. E-mail: ywxie@zju.edu.cn

**Abstract**

Superconducting diodes, which enable dissipationless supercurrent flow in one direction while blocking it in the reverse direction, are emerging as pivotal components for superconducting electronics. The development of editable superconducting diodes could unlock transformative applications, including dynamically reconfigurable quantum circuits that adapt to operational requirements. Here, we report the first observation of the superconducting diode effect (SDE) in $LaAlO_3/KTaO_3$ heterostructures—a two-dimensional oxide interface superconductor with exceptional tunability. We observe a strong SDE in Hall-bar (or strip-shaped) devices under perpendicular magnetic fields (< 15 Oe), with efficiencies above 40% and rectification signals exceeding 10 mV. Through conductive atomic force microscope lithography, we demonstrate reversible nanoscale editing of the SDE's polarity and efficiency by locally modifying the superconducting channel edges. This approach enables multiple nonvolatile configurations within a single device, realizing an editable superconducting diode. Our work establishes $LaAlO_3/KTaO_3$ as a platform for vortex-based nonreciprocal transport and provides a pathway toward designer quantum circuits with on-demand functionalities.

The nonreciprocal transport of charge carriers—exemplified by semiconductor *p*–*n* junctions—is the cornerstone of modern electronics [1]. In superconducting circuits, an analogous nonreciprocal supercurrent transport, known as the superconducting diode effect (SDE), has recently emerged as a critical functionality for quantum technologies [2]. Unlike classical diodes, the SDE requires simultaneous breaking of time-reversal and spatial inversion symmetries [2], a condition typically achieved through two mechanisms: (i) finite-momentum Cooper pairing (mediated by spin-orbit coupling [3] or Meissner screening [4]) or (ii) asymmetric vortex dynamics [5,6]. While early SDE signatures were observed in superconducting quantum interference devices (SQUIDs) [7,8], recent advances have expanded the material toolkit to include heterostructure multilayers [9,10], Josephson junctions [11–13], nanostructured vortex rectifiers [5,14,15], and thin films [16,17].

Two-dimensional oxide interface superconductors (2DOISs) [18–20], particularly $LaAlO_3/KTaO_3$ (LAO/KTO) heterostructures [19,20], offer unique advantages for realizing and controlling SDE. These systems naturally exhibit spatial inversion symmetry breaking due to their structural asymmetry [21–23] while simultaneously possessing the key criteria for both SDE mechanisms: the strong spin-orbit coupling from 5d Ta orbitals in KTO [24] enables possible finite-momentum Cooper pairing near the interface [25], while the extremely low superfluid density [26] and intrinsic two-dimensional nature [19,20] create ideal conditions for studying vortex dynamics and magnetic flux behavior [27,28]. Beyond these fundamental advantages, 2DOISs demonstrate exceptional tunability—their superconducting states can be well controlled through both global substrate gating [29,30] and local conductive atomic force microscope (cAFM) lithography [28,31–33]. This dual control capability opens new possibilities for post-fabrication SDE engineering, a feature that remains experimentally unexplored despite existing applications of superconducting diodes in full-wave rectification circuits [34,35].

In this work, we report a robust SDE in LAO/KTO microdevices under small perpendicular magnetic fields (< 15 Oe), observed in multiple device configurations including four-probe Hall-bar, eight-probe Hall-bar, and two-probe π-bar geometries. Our measurements reveal maximum rectification efficiencies of ~40% with AC output voltages reaching tens of

millivolts—significantly exceeding values reported for artificially engineered asymmetric nano-defect devices [6,14,15] and SQUID [36]. Furthermore, we demonstrate non-volatile modification of the SDE through cAFM lithography by nanoscale alteration of channel edge morphology. This tunability directly confirms the vortex edge asymmetry origin while establishing a practical approach for reconfigurable vortex-mediated SDEs.

## Sample and measurement setup

The LAO/KTO 2DOIS was formed by depositing LAO films on (111)-oriented KTO single-crystal substrates pre-patterned into Hall-bar configurations (see Supplemental Material for details [37]). As shown in Fig. 1(a), electrical transport measurements used a four-probe configuration with current (+*x*) and perpendicular magnetic field (+*z*) directions defined in Figs. 1(a) and 1(b). Current-voltage (*I-V*) characteristics were measured using triangular current sweeps: positive sweeps followed 0 to +*I* (0–p), +*I* to -*I* (p-0 and 0-n), and -*I* to 0 (n-0) (shown in Fig. 1(d) inset), while negative sweeps used the reverse sequence (0–n, n–p, p-0), both yielding equivalent results (Supplemental Material Fig. S1).

## Superconducting diode effect

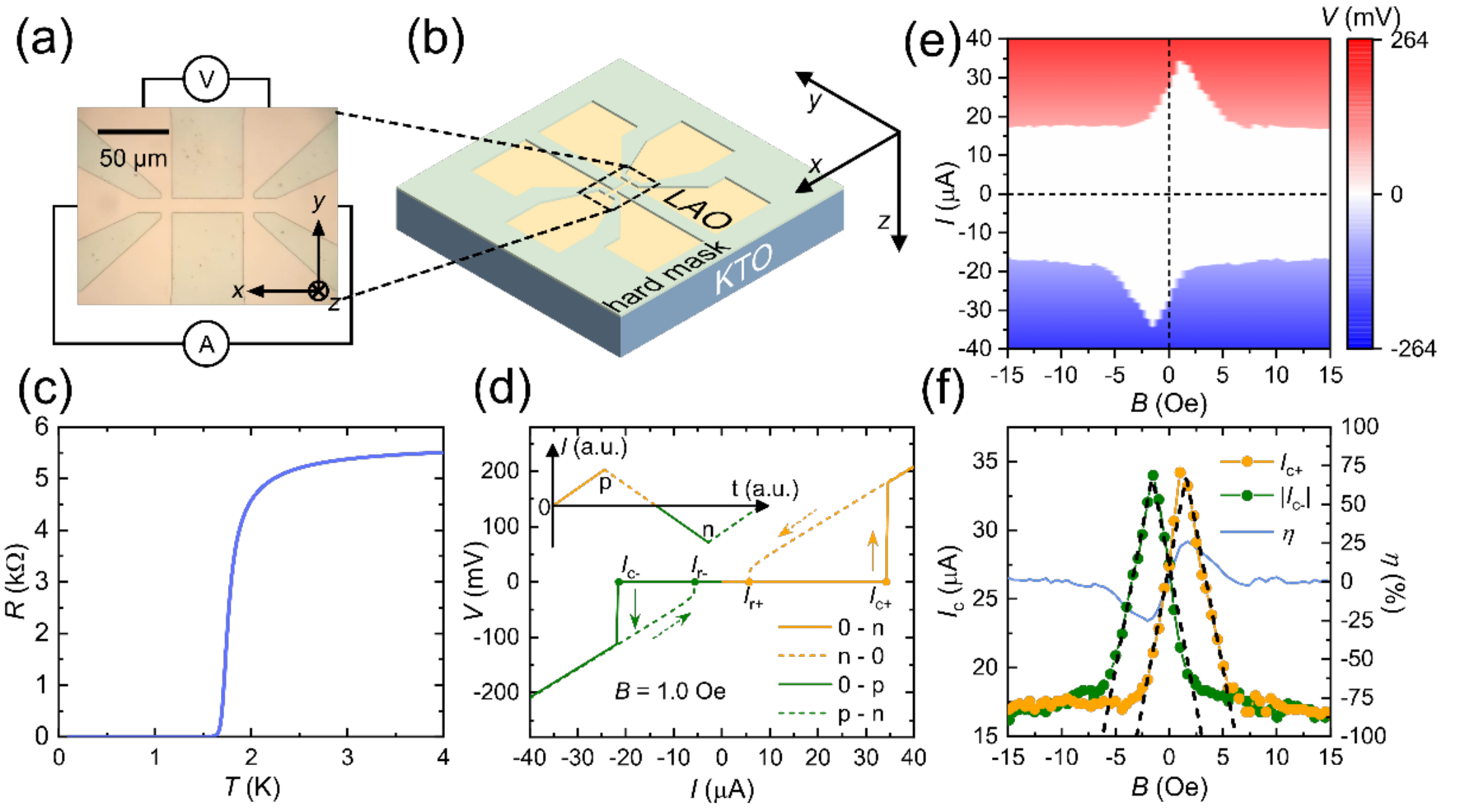

Fig. 1. Superconducting diode effect in a LAO/KTO Hall-bar device. Electrical transport characterization of Device A1 (10 μm × 50 μm four-probe Hall-bar geometry) at 100 mK under perpendicular magnetic fields. (a) Optical micrograph of the device. (b) Schematic with defined coordinate axes: current (*I*, +x) and magnetic field (*B*, +z). (c) Temperature-dependent resistance demonstrating the superconducting transition. (d) *I-V* characteristics at 1.0 Oe with current sweep sequence (inset). (e) Voltage ($V$) versus current ($I$) and field ($B$) color map, revealing nonreciprocal transport. (f) Field dependence of critical currents (left axis) and diode efficiency $\eta$ (right axis), with dashed lines showing fits to the vortex edge asymmetry model (parameters in Supplemental Material Table S1).

As shown in Fig. 1(c), Device A1 exhibits superconductivity with a midpoint transition temperature $T_c \approx 1.8$ K and reaches zero-resistance state at ~1.4 K. Figure 1(d) displays $I$-$V$ characteristics at 100 mK ($\ll T_c$) under a 1.0 Oe perpendicular field. The $I$-$V$ curve shows anticlockwise hysteresis attributed to Joule self-heating or flux-flow instability [5,27,38], with the critical current for positive direction ($I_{c+}$) significantly exceeding that for negative bias ($I_{c-}$), confirming SDE. The voltage map in Fig. 1(e), obtained by measuring $I$-$V$ characteristics across varying magnetic fields, reveals pronounced asymmetry around $B = 0$ that directly correlates with the $I_{c+}/I_{c-}$ disparity. Figure 1(f) shows the field dependence of $I_{c+}$ and $|I_{c-}|$, exhibiting an inverted "V" shape with peaks at finite fields. Crucially, the magnitudes interchanges upon field reversal: $I_{c+} > |I_{c-}|$ for $B > 0$, while $I_{c+} < |I_{c-}|$ for $B < 0$. The diode efficiency $\eta = (I_{c+} - |I_{c-}|)/(I_{c+} + |I_{c-}|)$ reaches ~26% at ~1.6 Oe (Fig. 1(f)).

Beyond the four-probe Hall-bar device (A1) in Fig. 1, we fabricated and characterized additional devices, including eight-probe Hall-bar and two-probe π-bar configurations, using identical methods. All devices exhibited SDE, though with sample-dependent variations in polarity and efficiency (Supplemental Material Figs. S2, S3, S4). The strongest effect was observed in a 3 μm × 90 μm four-probe Hall-bar device, achieving ~40% efficiency (Supplemental Material Fig. S4), corresponding to a critical current asymmetry ratio $|I_{c+}/I_{c-}| \approx 230\%$. These results confirm the SDE's ubiquity across our platform.

**Current rectification**

A fundamental function of diodes is rectification—converting alternating current (AC) to direct current (DC) [6,17,39,40]. We investigated the rectifying behavior of the SDE using both low-frequency square-wave (5 Hz) and high-frequency sine-wave AC (1 kHz). In Device C1 at $B = -5.9$ Oe (Fig. 2(a)), applying a square-wave current (bottom panel) produced three distinct response regimes in the measured voltage drop (top panel): (1) for ±8 μA ($|I_{applied}| < I_{c+}$ and $|I_{c-}|$), the device remained superconducting with zero voltage drop; (2) ±10 μA ($I_{c+} < |I_{applied}| < |I_{c-}$

|), clear half-wave rectification occurred—the channel remained superconducting at -10 μA but switched to the normal state at +10 μA; (3) For ±12 μA ($|I_{applied}| > I_{c+}$ and $|I_{c-}|$), the rectification vanished as the device entered the normal state for both current polarities.

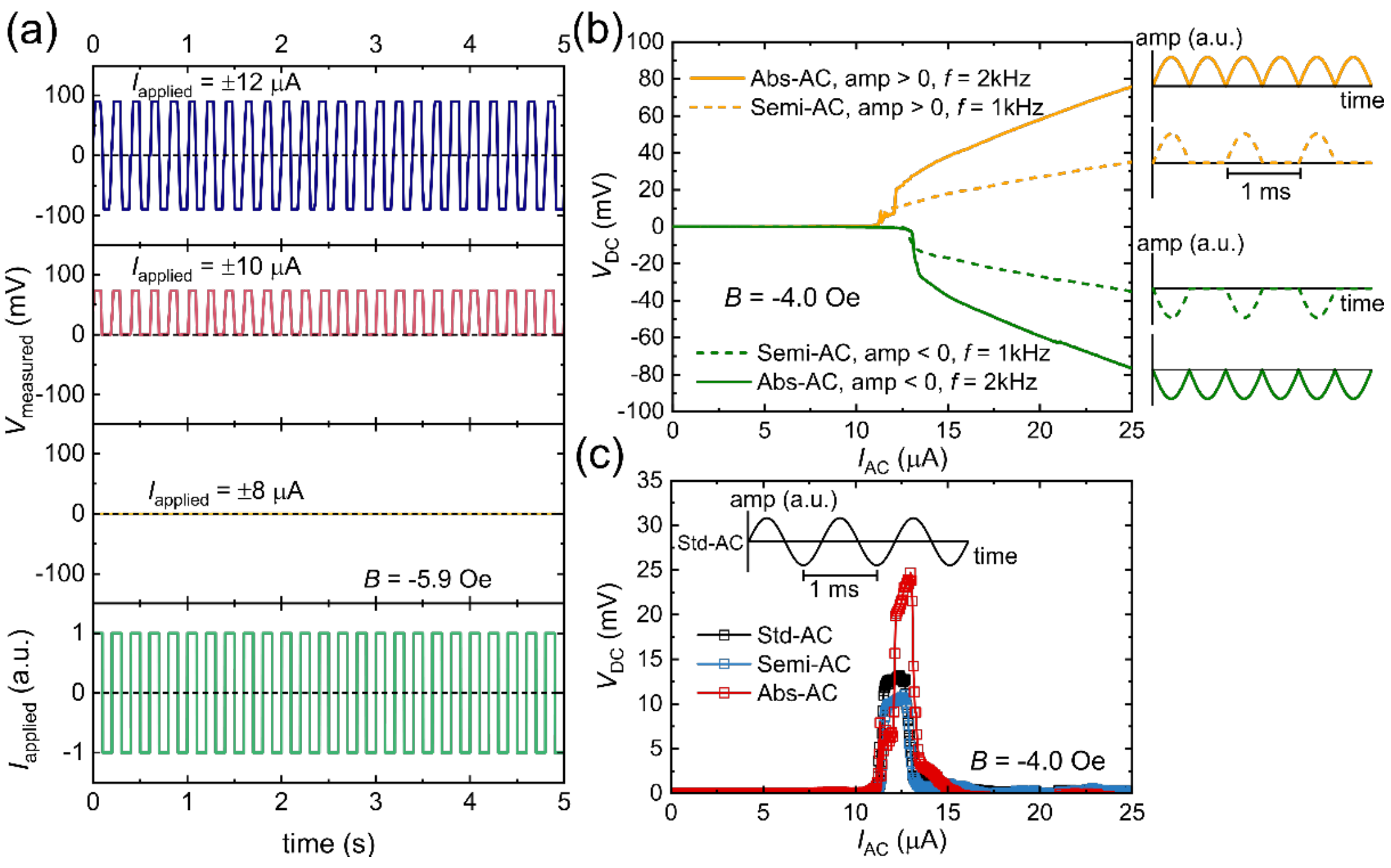

Fig.2. Current rectification in Device C1 at 300 mK. (a) Half-wave rectification response under $B$ = -5.9 Oe. (b) DC voltage response to Semi-AC and absolute-AC (Abs-AC) current inputs at $B$ = -4.0 Oe, with corresponding current waveforms shown in the right panel. (c) Comparison between calculated rectification signals from panel (b) and standard AC (Std-AC) experiment (black).

In addition to half-wave square-wave rectification, we employed the quasi-DC method proposed by Lyu et al. [6] to quantitatively analyze sine-wave rectification. As shown in Fig. 2(b) (right panel), we applied two current waveforms to the device: (1) a fully rectified 2 kHz sine wave AC (Abs-AC) and (2) a half-rectified 1 kHz sine wave AC (Semi-AC), while measuring the resulting DC voltage (left panel). These quasi-DC currents maintain sine-wave oscillations but with fixed polarities. At $B$ = -4.0 Oe, the device exhibited polarity-dependent transition current shifts, consistent with SDE behavior (Supplemental Material Fig. S3(f)). By superimposing the positive and negative quasi-DC responses, we obtained rectification signals (Fig. 2(c)) that matched standard 1 kHz AC measurements (black curve), with Semi-AC rectification amplitude equaling the Std-AC response (half of Abs-AC), providing definitive SDE evidence. Notably, the Std-AC rectification voltage (~tens of millivolts) significantly surpasses both flux-quantum diode effects (by $10^4\times$) [15] and conformal-mapped devices (by ~2×) [6]. This giant rectification voltage arises from LAO/KTO's characteristically high normal-state resistivity and can be further optimized through geometric control of either the channel dimensions or superconducting layer thickness [6].

**Mechanism of the SDE**

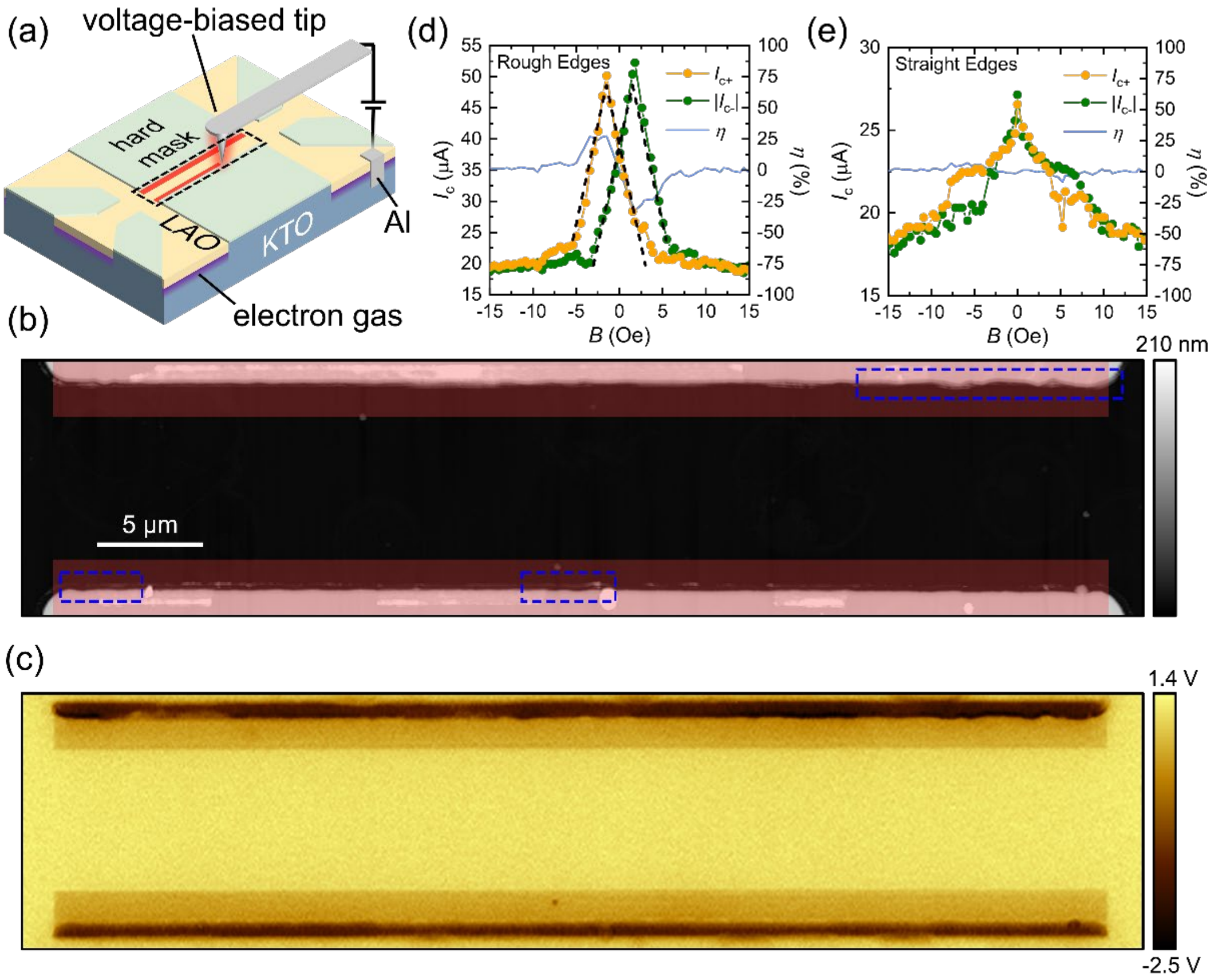

Fig. 3. Verification of vortex edge asymmetry-induced SDE in Device A4 (10 μm × 50 μm Hall bar) via cAFM lithography. (a) Schematic of straight-edge channel fabrication using cAFM lithography. (b) Surface morphology showing LAO/KTO channel (dashed lines in (a)), where photolithography-produced edges appear rough (bright: hardmask; dark: LAO), while cAFM patterning creates straight edges by insulating red areas. (c) Surface potential image of the straight-edge channel (original channel in Supplemental Material Fig. S5). (d)-(e) SDE characteristics before (d) and after (e) cAFM processing, with dashed lines showing fits to the vortex edge asymmetry model (parameters in Supplemental Material Table S1).

First, we rule out finite-momentum Cooper pairing scenarios. In such cases—whether in 2D superconductors [41,42] or Josephson junctions [4,13]—the magnetic field is applied along the *y*-direction (parallel to the film plane), creating nonreciprocal critical currents in the *x*-direction.

By contrast, our work employs a distinct geometry with the field oriented along $z$ (perpendicular to the interface).

Instead, we attribute the SDE in our LAO/KTO devices to vortex edge asymmetry [5,43], where fabrication-induced defects (e.g., wave-like structures in Fig. 3(b) blue dashed boxes), create unequal surface barriers for vortex entry at opposite channel edges. This asymmetry breaks the mirror symmetry in the $x$-$z$ plane, yielding different critical current densities ($j_c$) at each edge. Under perpendicular magnetic fields, the Meissner effect generates counterflowing screening currents that combine with this inherent $j_c$ asymmetry to produce nonreciprocal critical currents ($I_{c+} \neq |I_{c-}|$). The vortex edge asymmetry model quantitatively explains our results, with field-dependent critical currents following [5]:

$$I_{c+} = \begin{cases} S\left(j_c + \delta j_c - aB\right), B \geq \dfrac{\delta j_c}{2a} \\ S\left(j_c + aB\right), B < \dfrac{\delta j_c}{2a} \end{cases} \tag{1}$$

$$I_{c-} = \begin{cases} S\left(j_c - aB\right), B \geq -\dfrac{\delta j_c}{2a} \\ S\left(j_c + \delta j_c + aB\right), B < -\dfrac{\delta j_c}{2a} \end{cases} \tag{2}$$

where $S$ is the channel cross-section, $\delta j_c$ the edge asymmetry in $j_c$, and $a$ the Meissner response coefficient. This model fits all data (Figs. 1(f), 3(d), and Supplemental Material Figs. S2(c), S3(b), S3(d), S3(f), S4(c)) with parameters in Supplemental Material Table S1 [37].

We conclusively verified this mechanism through cAFM lithography, which enables local and non-volatile control of LAO/KTO interface conductivity at the nanoscale via voltage-biased tip scanning (Fig. 3(a)) —negative bias drives the interface toward insulating, while positive bias toward conducting [28,33]. By transforming rough-edged channels into straight-edged channel configurations (Figs. 3(a)-3(c)), we eliminated the SDE, observing near-overlapping $I_{c+}$ and $|I_{c-}|$ near zero field with reduced diode efficiency (Fig. 3(e) vs. Fig. 3(d)). This controlled modification of edge morphology directly links SDE tuning to vortex barrier engineering, while definitively excluding finite-momentum pairing contributions.

**Nanoscale engineering of SDE through vortex edge control**

The near-disappearance of SDE in the straight-edged channel (Fig. 3(e)) not only confirms its vortex-edge origin but also demonstrates the effect's remarkable sensitivity to edge geometry. Using Device A2 as a platform, we systematically engineered different edge configurations: the original rough-edged channel (Fig. 4(a)) exhibited a characteristic inverted "V" shaped field-dependent $I_c$ with a ~11% maximum diode efficiency, while channels with comb-patterned bottom edges (~2 μm, ~5 μm, and ~7 μm tooth lengths: Fig. 4(c) and Supplemental Material Fig. S6) showed enhanced SDE, peaking at ~17% efficiency for the ~5 μm pattern (Fig. 4(d)). Further modification to create a straight bottom edge with rough top edge (Fig. 4(e)) yielded ~30% efficiency but with reversed polarity ($I_{c+} < |I_{c-}|$ for $B > 0$), consistent with the vortex edge asymmetry scenario [5,44] where straight edges develop higher surface barriers. Finally, refining both edges to straight (Fig. 4(g)) significantly weakened SDE, matching the control experiment (Fig. 3(e)). This programmable evolution of SDE through cAFM lithography demonstrates a reconfigurable quantum circuit element where superconducting edge morphology can be precisely tailored to achieve desired diode characteristics.

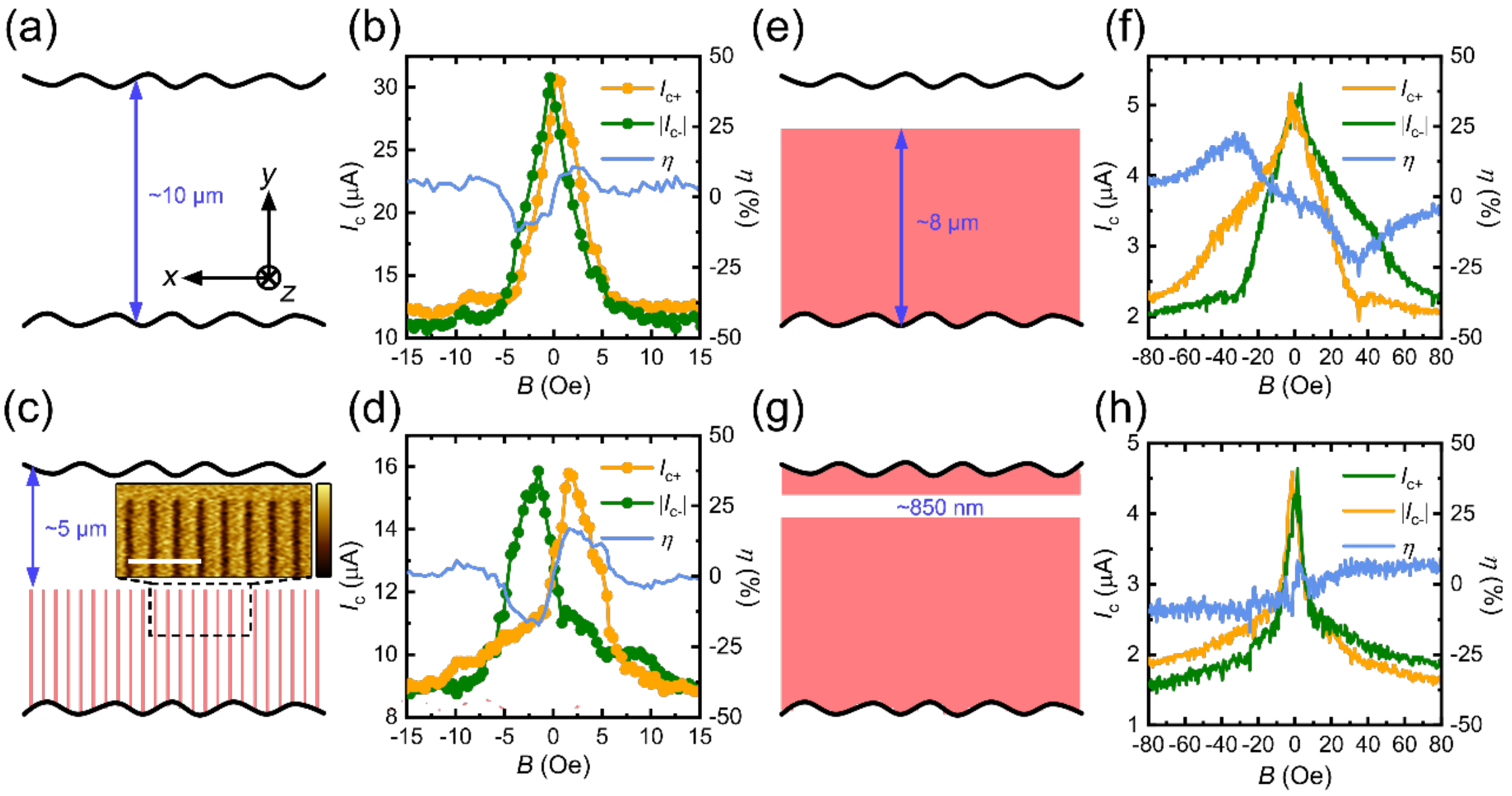

Fig. 4. Editable SDE modification in Device A2 via cAFM lithography. (a) Schematic of the as-fabricated channel with rough edges. (b) SDE characteristics corresponding to (a). (c) Engineered channel with comb-patterned bottom edge (~5 μm tooth length) and original rough top edge (inset: surface potential image; scale bar: 2 μm; color bar: 0.5 V (dark) to 1.2 V (bright); full image in Supplemental Material Fig. S7(b)). (d) SDE response for edge configuration in (c). (e) Channel with straight bottom edge and rough top edge. (f) SDE for configuration (e). (g) Fully engineered straight-edge channel. (h) SDE characteristics for (g).

## Conclusion

In summary, we observed asymmetric critical currents ($I_{c+} \neq |I_{c-}|$) in LAO/KTO Hall-bar devices under small out-of-plane fields (< 15 Oe) and demonstrated SDE through AC rectification measurements. The effect originates from fabrication-induced edge defects that create unequal vortex entry conditions, as described by the vortex edge asymmetry model. Crucially, we achieved non-volatile control of both SDE polarity and efficiency through nanoscale edge engineering via cAFM lithography. These findings establish oxide interfaces as a versatile platform for both fundamental studies of vortex-mediated rectification and the development of reconfigurable quantum circuits.

**Acknowledgments:** This work was supported by the National Natural Science Foundation of China (12534005, 12325402) and the National Key R&D Program of China (2023YFA1406400).

**Author contributions:** Y.X. supervised the project. Y.W. and W.P. grew LAO/KTO samples and conducted transport measurements. Y.W. performed the cAFM lithograph. M.Z. contributed to the experimental setup. Y.W. and Y.X. analyzed the data and wrote the manuscript with inputs from all authors.

**Competing interests:** The authors declare no competing interests.

**Additional information:** Data for all graphs presented in this paper are available upon reasonable request (Y. X.).

# Supplementary Material for

## $KTaO_3$-based editable superconducting diode

Yishuai Wang[#], Wenze Pan[#], Meng Zhang and Yanwu Xie*

Correspondence: ywxie@zju.edu.cn

**This file includes:**

**Materials and Methods.**

1. Device fabrication

**Hall-bar device fabrication:** Using standard optical lithography and lift-off techniques, we patterned clean KTO (111) surface into Hall-bar geometries by depositing an amorphous $AlO_x$ (a-$AlO_x$) hard mask on non-device areas (Fig. 1(b)). A 6-10 nm thick amorphous LAO (a-LAO) layer was then deposited, forming the superconducting LAO/KTO interface exclusively in unmasked Hall-bar regions while a-$AlO_x$-covered areas remained highly insulating.

**Film deposition:** Both $AlO_x$ and LAO films were deposited by pulsed laser deposition using a 248-nm KrF excimer laser. $AlO_x$ hard masks were deposited at room temperature with 2 $Jcm^{-2}$ laser fluence at 3-4 Hz repetition rate, followed by oxygen annealing to ensure highly insulating $AlO_x$/KTO interfaces. LAO films were grown at 300 °C in a mixed atmosphere ($1\times10^{-5}$ mbar $O_2$ and $1\times10^{-7}$ mbar $H_2O$ vapor) using ~0.5 $Jcm^{-2}$ fluence and 10 Hz repetition rate, with subsequent cooling to room temperature in the growth atmosphere.

**cAFM lithography:** Conductive atomic force microscopy (cAFM) lithography was performed using a commercial AFM system (Park NX10) under ambient conditions. Pt/Cr-coated tips (Multi75E-G, Budget Sensors) were used, with key parameters (voltage bias, contact force, tip velocity) detailed in Supplemental Material Table S2.

2. Surface potential measurements

Surface potential characterization was performed using single-pass Kelvin probe force microscopy (KPFM) [45] on the same AFM system (Park NX10) with identical Pt/Cr-coated conductive tips (Multi75E-G, Budget Sensors), under ambient conditions. The typical KPFM parameters are $V_{ac}$ of 0.8 V (peak-to-peak), $f_{resonace}$ of 17 kHz, and scan rate of 0.2 Hz.

3. Electrical contacts and transport measurements

**Contacts.** The conducting LAO/KTO interface was contacted via ultrasonic Al wire bonding.

**Transport measurements.** All measurements were performed in a Quantum Design PPMS dilution refrigerator using the standard four-probe method in Hall-bar geometries. Ohmic contacts were made through ultrasonic Al wire bonding. To eliminate photoconduction effects, samples were kept in dark for ≥5 hours prior to measurements. Device resistance and $I$-$V$ characteristics were measured using the PPMS Resistivity Option and Electrical Transport Option. Rectification measurements employed: (i) low-frequency (5 Hz) square waves (PPMS Resistivity Option) and (ii) high-frequency (1 kHz and 2 kHz) sine waves (Keithley 6221 current source), with DC output detected by a Keithley 2182A nanovoltmeter.

4. Zero field corrections

The magnetic field was applied using a superconducting magnet, where trapped flux complicates the determination of the true zero-field point [46]. We established the zero-field reference through a three-step protocol:

(1) Measuring magnetoresistance slightly above the zero-resistance temperature (red curve, Supplemental Material Fig. S8(c) upper panel);
(2) Cooling to low temperature for SDE measurements (*I-V* curves at various fields, Supplemental Material Fig. S8(c) lower panel);
(3) Repeating the magnetoresistance measurement after warming (blue curve, Supplemental Material Fig. S8(c) upper panel).

The identical magnetoresistance curves before and after temperature cycling confirm stable field conditions. The symmetric magnetoresistance response provides a reliable zero-field reference, with verification showing equal critical current ($I_{c+} = |I_{c-}|$) and zero efficiency ($\eta = 0$) at zero field, while demonstrating unequal critical current ($I_{c+} \neq |I_{c-}|$) and non-zero efficiency ($\eta \neq 0$) at finite fields, as shown in Supplemental Material Fig. S8(d). Throughout this work, zero-field was defined as either the magnetoresistance minimum or the point where diode efficiency vanishes (Supplemental Material Fig. S8).

5. Vortex edge asymmetry model

The model assumes: (1) linear response of Meissner screening current density at channel edges to external field $B$ with slope $a$, and (2) a $\delta j_c$ difference in critical current density between upper and lower edges. The field-dependent critical currents are given by [5]:

$$I_{c+} = \begin{cases} S\left(j_c + \delta j_c - aB\right), B \geq \dfrac{\delta j_c}{2a} \\ S\left(j_c + aB\right), B < \dfrac{\delta j_c}{2a} \end{cases} \tag{1}$$

$$I_{c-} = \begin{cases} S\left(j_c - aB\right), B \geq -\dfrac{\delta j_c}{2a} \\ S\left(j_c + \delta j_c + aB\right), B < -\dfrac{\delta j_c}{2a} \end{cases} \tag{2}$$

where $S$ is the channel cross-section and $j_c$ the critical current density at the bottom edge. Our experimental data (Figs. 1(f) and 3(d); Supplemental Material Figs. S2(c), S3(b), S3(d), S3(f)) are well fitted by this model, with parameters summarized in Supplemental Material Table S1.

TABLE S1. Parameters obtained by fitting field-dependent critical currents using the vortex edge asymmetry model.

| | $Sj_c$ | $S\delta j_c$ | $Sa$ |
|---|---|---|---|
| Device A1 | 27.44 μA | 12.94 μA | 4.09 μA/Oe |
| Device A2 | 28.19 μA | 3.63 μA | 3.63 μA/Oe |
| Device A3 | 45.80 μA | -9.03 μA | 5.10 μA/Oe |
| Device A4 | 58.30 μA | -19.32 μA | 6.40 μA/Oe |
| Device A5 | 12.91 μA | -3.83 μA | 0.51 μA/Oe |
| Device B1 | 6.52 μA | 1.29 μA | 1.75 μA/Oe |
| Device C1 | 16.42 μA | 2.81 μA | 1.28 μA/Oe |

TABLE S2. The cAFM lithography parameters.

| | *Contact force* | *Tip velocity* | *Voltage bias* |
|---|---|---|---|
| Fig. 3(b) | 10 nN | 2.0 µm/s | -9 V |
| Fig. S6(a) | 10 nN | 1.6 µm/s | -9 V |
| Fig. 4(c) | 10 nN | 1.6 µm/s | -9 V |
| Fig. S6(c) | 10 nN | 1.6 µm/s | -9 V |
| Fig. 4(e) | 10 nN | 10.3 µm/s | -10 V |
| Fig. 4(g) | 10 nN | 10.3 µm/s | -10 V |

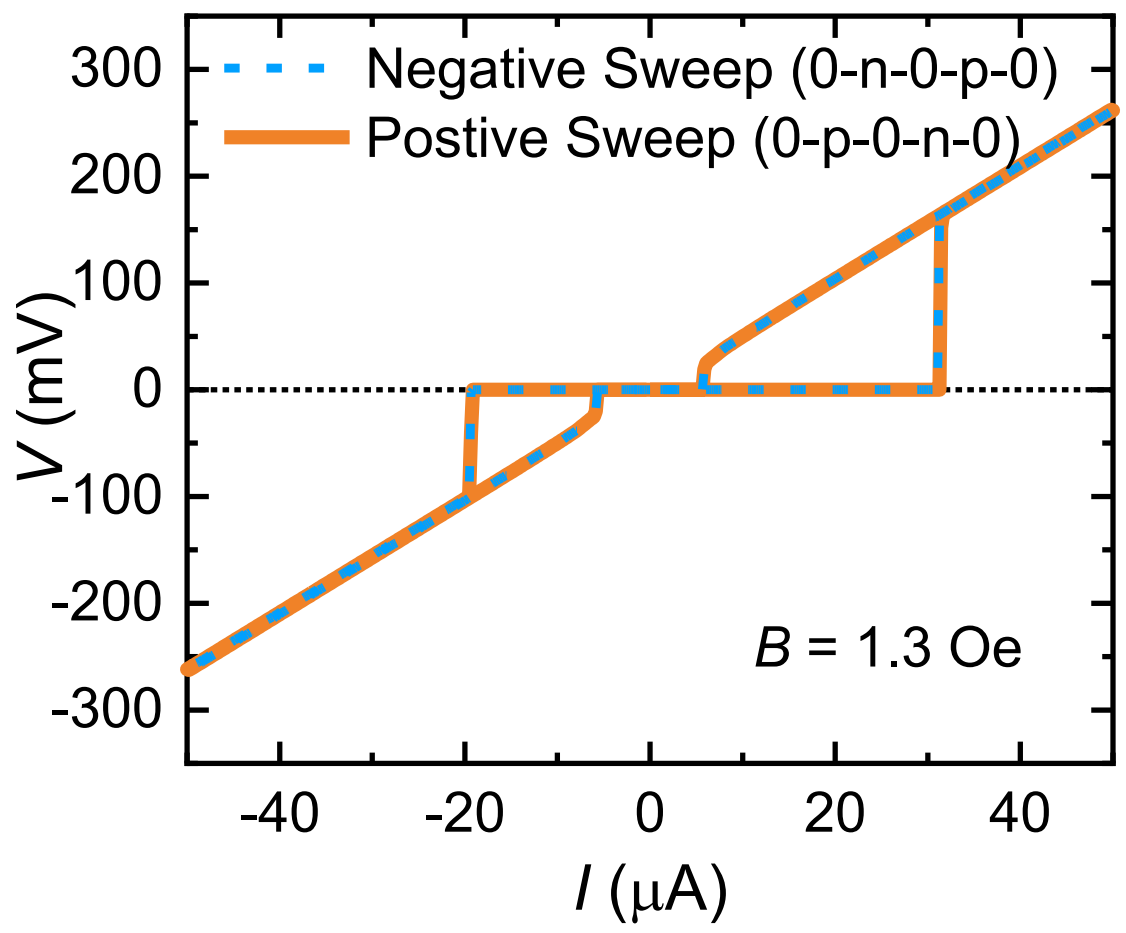

Fig. S1. Current-voltage characteristics under opposite sweep directions. *I*-*V* curves measured with positive current sweep (0→+*I*→-*I*→0, orange solid line) and negative sweep (0→-*I*→+*I*→0, blue dashed line), following the sweep sequences defined in the main text.

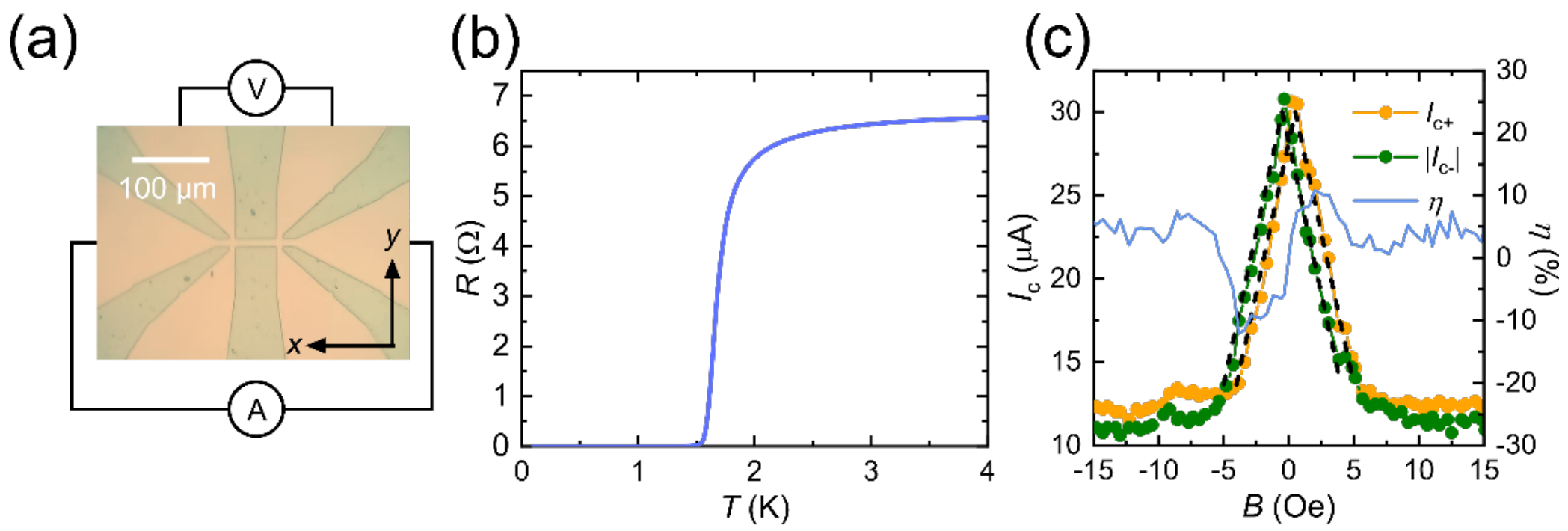

Fig. S2. Superconducting diode effect (SDE) in Device A2 (10 μm × 50 μm Hall-bar). (a) Optical micrograph showing the device geometry. (b) Temperature-dependent resistance demonstrating the superconducting transition. (c) Magnetic field dependence of critical currents and diode efficiency, revealing characteristic asymmetric behavior.

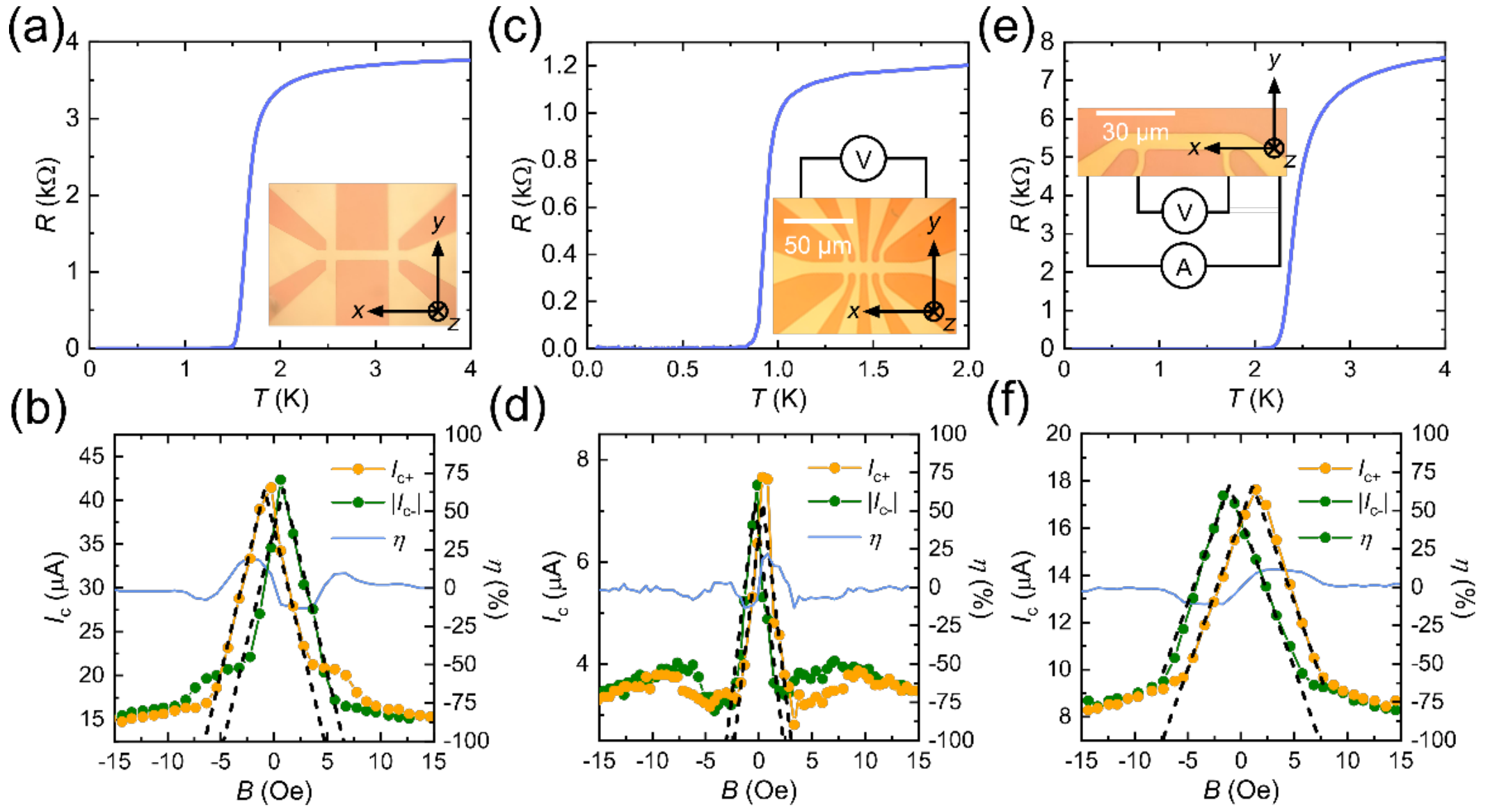

Fig. S3. SDE effect across multiple device geometries. Characterization of three devices: Device A3 (10 μm × 50 μm four-probe Hall-bar), Device B1 (8 μm × 20 μm eight-probe Hall-bar), and Device C1 (8 μm × 30 μm two-probe π-bar). (a)(c)(e) Temperature-dependent resistance showing superconducting transitions for (a) A3, (c) B1, and (e) C1, with optical micrographs in insets. (b)(d)(f) Corresponding field-dependent critical currents and diode efficiency at (b) 100 mK (A3), and (d)(f) 300 mK (B1, C1), with dashed lines showing fits to the vortex edge asymmetry model (parameters in Supplementary Material Table S1).

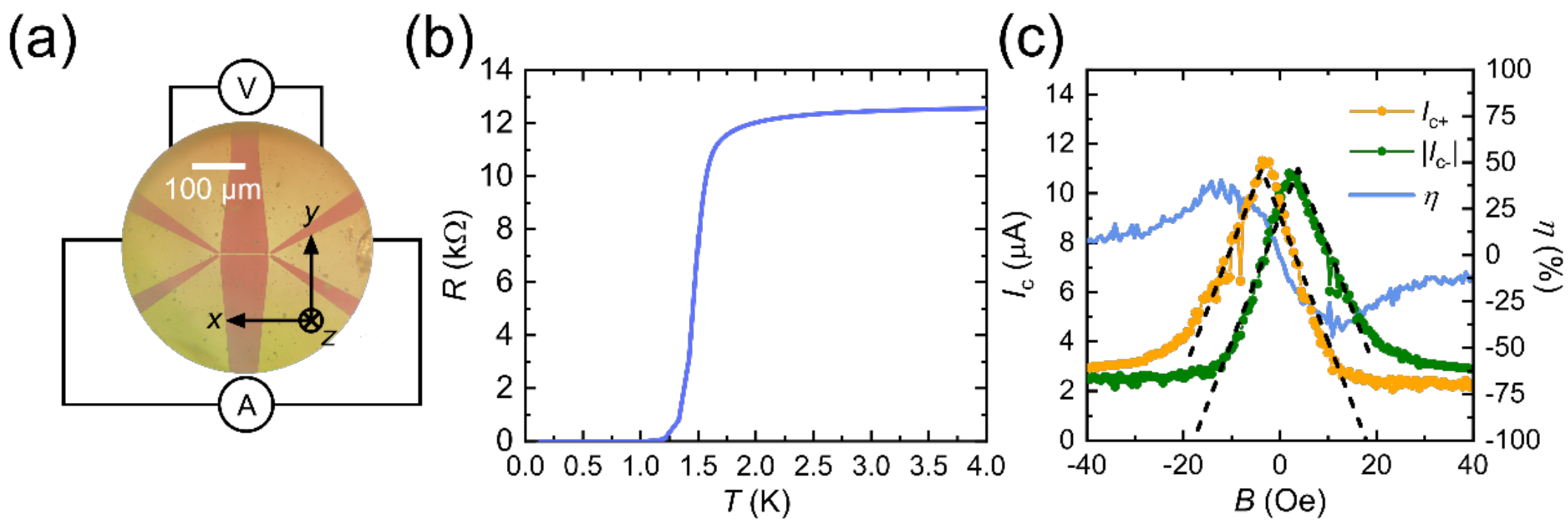

Fig. S4. Enhanced SDE in Device A5 (3 μm × 90 μm Hall-bar). (a) Optical micrography showing the elongated channel geometry. (b) Temperature-dependent resistance demonstrating a superconducting transition. (c) Field-dependence of critical currents and diode efficiency, revealing enhanced asymmetry compared to standard devices.

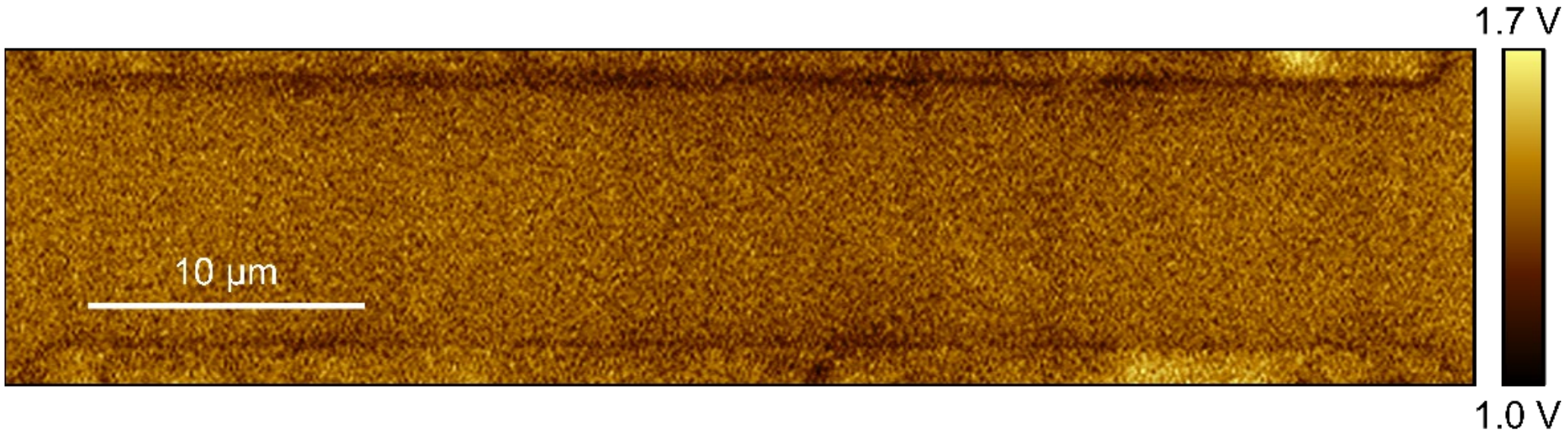

Fig. S5. Surface potential mapping of the as-fabricated channel in Device A4. KPFM image showing the intrinsic potential distribution across the unmodified superconducting channel prior to cAFM lithography processing.

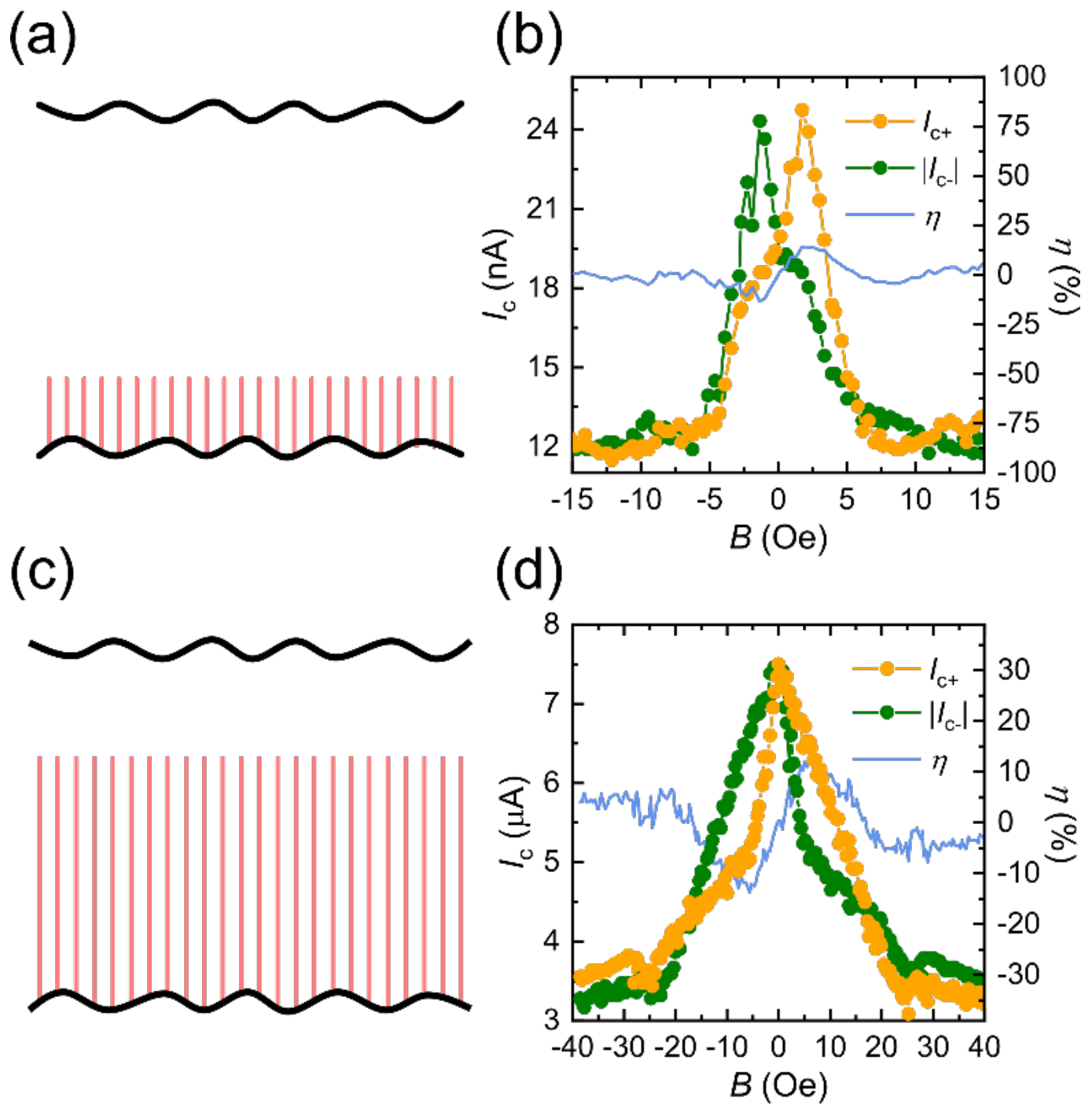

Fig. S6. Engineered comb-like edge structures and their SDE response. (a) Schematic of channel with ~2 μm comb-patterned bottom edge (rough top edge retained); surface potential map shown in Supplemental Material Fig. S7(a). (b) Corresponding SDE characteristics. (c) Schematic of channel with ~7 μm comb-patterned bottom edge (rough top edge retained); surface potential map shown in Supplemental Material Fig. S7(c). (d) Corresponding SDE characteristics.

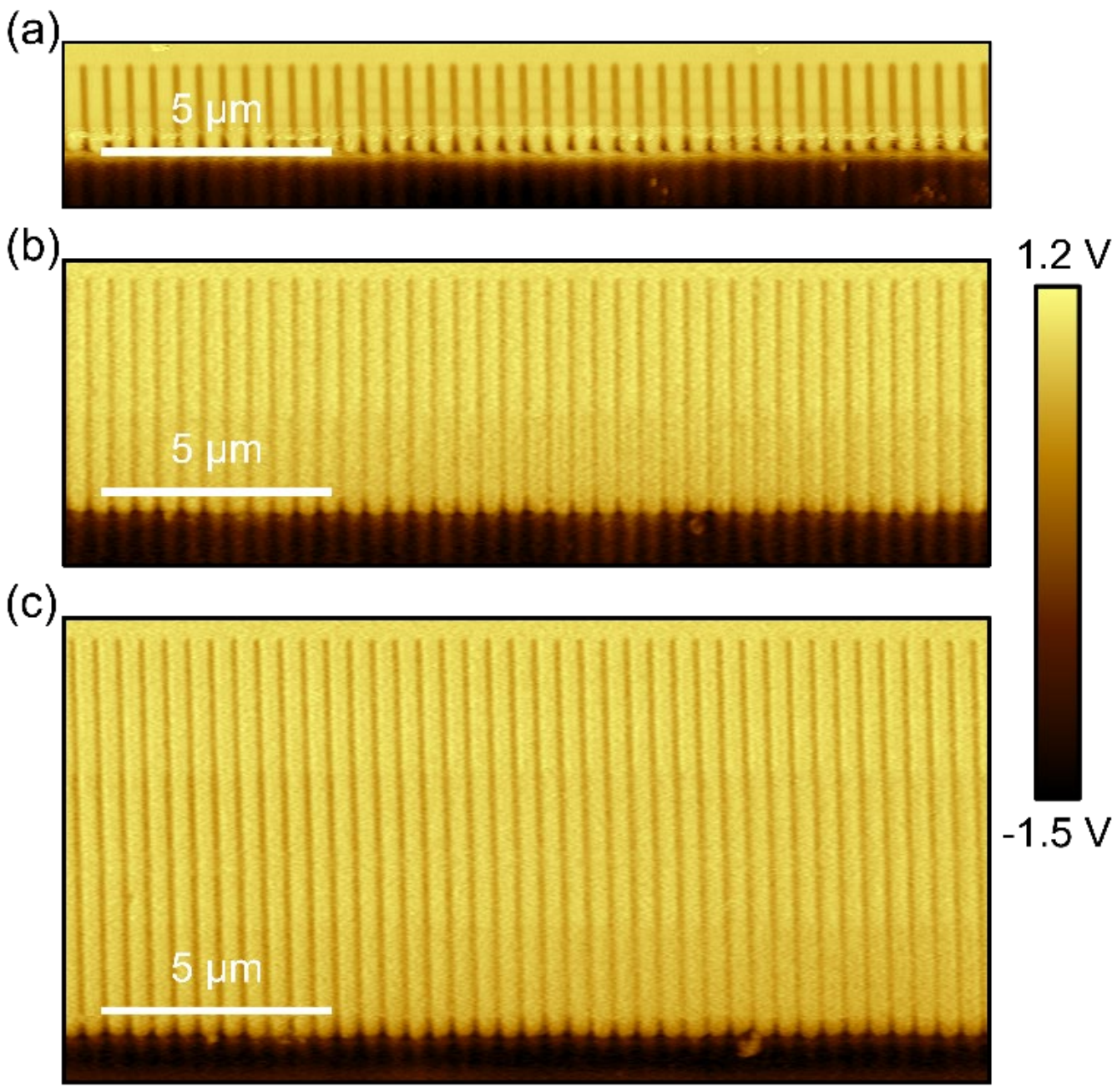

Fig. S7. Surface potential mapping of engineered edge structures. (a)Surface potential of ~2 μm comb-patterned (Supplemental Material Fig. S6(a) configuration). (b) Surface potential of ~5 μm comb-patterned (Fig. 4(c) configuration). (c) Surface potential of ~7 μm comb-patterned (Supplemental Material Fig. S6(c) configuration). All images reveal potential variations at patterned edges, correlating with SDE modifications in corresponding transport measurements.

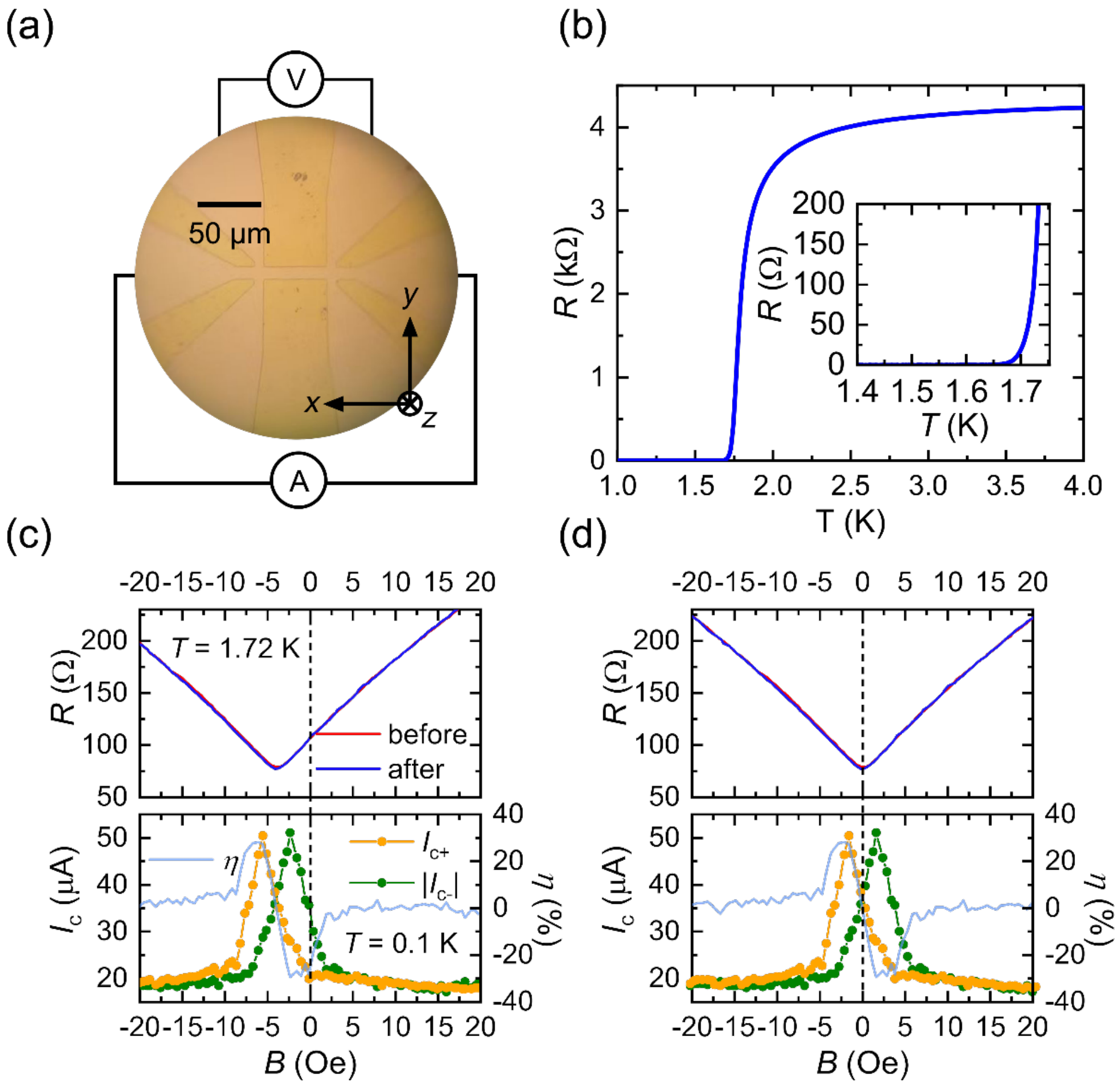

Fig. S8. Zero magnetic field calibration protocol for Device A4. (a) Optical micrograph of the measurement device. (b) Temperature-dependent resistance with inset showing detailed superconducting transition. (c) Upper panel: Magnetoresistances at 1.72 K (within superconducting transition regime) before (red) and after (blue) field-dependent $I_c$ measurements; Lower panel: Corresponding critical currents and diode efficiency versus field. (d) Calibrated field dependence of resistance, critical currents, and diode efficiency, with zero-field defined as the magnetoresistance minimum (upper panel reference).